\documentclass[reqno,12pt]{amsart}
\usepackage{fullpage}
\usepackage{mathtools}
\usepackage{amsfonts}
\usepackage{amssymb}
\usepackage{amsmath}
\usepackage{xcolor}

\def\D{{\mathcal D}}
\def\H{{\mathcal H}}
\def\E{{\mathcal E}}

\def\inv{{}^{-1}}

\def\sgn{{\rm sgn}}
\def\Rnum{\mathbb{R}}

\def\loc{{\rm loc}}
\def\const{\text{const}}

\def\Parder#1#2{\frac{\partial #1}{\partial #2}}

\newtheorem{prop}{Proposition}
\newtheorem{thm}{Theorem}

\newtheorem{lem}{Lemma}

\def\Ref#1{Ref.~\cite{#1}}
\def\secref#1{Sec.~\ref{#1}}

\def\scrpt#1{$\scriptstyle {#1}$}

\begin{document}
\allowdisplaybreaks[3]

\title{Hamiltonian structure of peakons\\ as weak solutions for the modified \\ Camassa-Holm equation}

\author{
Stephen C. Anco${}^1$
\\\lowercase{\scshape{ and }}\\
Daniel Kraus${}^2$
\\
\\\lowercase{\scshape{
${}^1$Department of Mathematics and Statistics\\
Brock University\\
St. Catharines, ON L\scrpt2S\scrpt3A\scrpt1, Canada}}
\\\lowercase{\scshape{
${}^2$Department of Mathematics\\
SUNY Oswego\\
Oswego, NY 13126}}
}

\thanks{sanco@brocku.ca, daniel.kraus@oswego.edu.}

\begin{abstract}
The modified Camassa-Holm (mCH) equation is a bi-Hamiltonian system 
possessing $N$-peakon weak solutions, for all $N\geq 1$, 
in the setting of an integral formulation 
which is used in analysis for studying local well-posedness, global existence, and wave breaking for non-peakon solutions. 
Unlike the original Camassa-Holm equation, 
the two Hamiltonians of the mCH equation do not reduce to conserved integrals 
(constants of motion) for $2$-peakon weak solutions. 
This perplexing situation is addressed here by finding 
an explicit conserved integral for $N$-peakon weak solutions for all $N\geq 2$. 
When $N$ is even, 
the conserved integral is shown to provide a Hamiltonian structure 
with the use of a natural Poisson bracket that arises from reduction of 
one of the Hamiltonian structures of the mCH equation. 
But when $N$ is odd, 
the Hamiltonian equations of motion arising from the conserved integral using this Poisson bracket 
are found to differ from the dynamical equations for the mCH $N$-peakon weak solutions. 
Moreover, the lack of conservation of the two Hamiltonians of the mCH equation 
when they are reduced to $2$-peakon weak solutions is shown to extend to $N$-peakon weak solutions for all $N\geq 2$. 
The connection between this loss of integrability structure 
and related work by Chang and Szmigielski on the Lax pair for the mCH equation is discussed. 
\end{abstract}

\maketitle

\section{Introduction}

Peakons are peaked travelling waves of the form $u(x,t)=a\exp(-|x-ct|)$
which arise as weak (non-smooth) solutions to nonlinear dispersive wave equations 
\begin{equation}\label{fg-fam}
m_t + f(u,u_x)m +(g(u,u_x)m)_x=0, 
\quad 
m=u-u_{xx}
\end{equation} 
where the relation between the amplitude $a$ and the wave speed $c$ of a peakon
is determined by the form of the nonlinearities $f$ and $g$ in a given wave equation. 
Weak solutions satisfy a standard integral formulation of these wave equations \eqref{fg-fam} given by \cite{RecAnc}
\begin{equation}\label{fg-fam-weak}
0  = \int_{0}^{\infty}\int_{-\infty}^{\infty} \big( 
\psi( u_t + f u + F_u u_x) 
-\psi_x( g u + G_u u_x -F) 
- \psi_{xx} (G  +u_t)
\big) \, dx \, dt 
\end{equation}
with $\psi(x,t)$ being a test function. 
This formulation is used in analysis for studying local well-posedness, global existence, and wave breaking for solutions of the Cauchy problem
\cite{ConStr,ConMol2000,EscLiuYin,GuiLiuTian,GuiLiuOlvQu}. 
Moreover, it regularizes product of distributions that would otherwise 
be undefined a priori for peakon solutions. 

Recent work \cite{RecAnc} has shown that every wave equation in the family \eqref{fg-fam},
with $f$ and $g$ being smooth functions of $u$ and $u_x$, 
possesses $N$-peakon weak solutions 
\begin{equation}\label{Npeakon}
u(x,t)=\sum_{i=1}^{N} a_i(t) \exp(-|x-x_i(t)|)
\end{equation}
whose time-dependent amplitudes $a_i(t)$ and positions $x_i(t)$
satisfy a nonlinear dynamical system consisting of $2N$ coupled ODEs 
\begin{equation}\label{fg-peakonsys}
\dot a_i = \tfrac{1}{2} [F(u,u_x)]_{x_i},
\quad
\dot x_i = -\tfrac{1}{2} [G(u,u_x)]_{x_i}/a_i
\end{equation}
in terms of $F(u,u_x)=\int f(u,u_x)\,d u_x$ and $G(u,u_x)=\int g(u,u_x)\,d u_x$, 
where square brackets denote the jump at a discontinuity.

Peakons were first found for the Camassa-Holm (CH) equation
\begin{equation}\label{CH}
m_t + 2u_xm +um_x =0
\end{equation}
which arises \cite{CamHol,CamHolHym} from the theory of shallow water waves
and is connected to the Korteveg-de Vries (KdV) equation  $u_t +uu_x +u_{xxx}=0$ 
by a reciprocal transformation \cite{Fuc}. 
Its $1$-peakon solutions have the speed-amplitude relation $c=a$,
while the dynamical system for its $N$-peakon solutions for all $N=1,2,\ldots$ 
has the structure \cite{CamHol} of a Hamiltonian system with a canonical Poisson bracket, 
where the Hamiltonian is given by the energy expression 
$E=\tfrac{1}{2}\sum_{i,j=1}^{N} a_i a_j \exp(-|x_i-x_j|)$. 
Moreover, 
this Hamiltonian peakon system is integrable in the Liouville sense of 
having $N$ commuting conserved integrals. 
Its integrability is connected to the property that 
the CH equation itself is integrable \cite{CamHol,FucFok} 
in the sense of possessing a Lax pair, a recursion operator, 
and a bi-Hamiltonian structure. 
In particular, 
one of the two Hamiltonian structures of the CH equation 
has a reduction \cite{HolHon} to peakon weak solutions, 
which yields the Hamiltonian structure of the $N$-peakon dynamical system. 

The nonlinearities in the both the KdV equation and the CH equation are quadratic. 
When cubic nonlinearities are considered, 
a natural counterpart of this pair of integrable equations is given by 
the modified KdV (mKdV) equation $u_t +u^2u_x +u_{xxx}=0$
and the modified CH (mCH) equation 
\begin{equation}\label{mCH}
m_t + ((u^2-u_x^2)m)_x =0
\end{equation}
which is also known as the FORQ equation. 
This peakon equation is connected to the mKdV equation by a reciprocal transformation \cite{Fok95a,Fok95b}
and is integrable \cite{OlvRos,Qia} in the same sense as the CH equation. 

There are, however, some very interesting differences 
when the peakons of the mCH equation and the CH equation are compared. 
The $1$-peakon solutions of the mCH equation have the speed-amplitude relation 
$c=\tfrac{2}{3}a^2$,
and consequently they are uni-directional, 
whereas the $1$-peakon solutions of the CH equation are bi-directional. 
In the dynamical system for the $N$-peakon solutions of the mCH equation,
\begin{equation}\label{mCHpeakonsys}
\dot a_i=0,
\quad
\dot x_i = [(\tfrac{1}{6} u_x^2-\tfrac{1}{2} u^2)u_x]_{x_i}/a_i, 
\end{equation}
all of the amplitudes $a_i$ are constant, 
and thereby this system cannot admit a canonical Hamiltonian structure. 
There is a non-canonical Poisson bracket \cite{HolHon} 
\begin{equation}\label{PB}
\{ a_i, a_j \} =0,
\quad
\{ a_i, x_j \} = - \{ x_j, a_i \} =0,
\quad
\{ x_i, x_j \}= \tfrac{1}{2}\sgn(x_i-x_j) 
\end{equation}
which arises from a reduction of one of the two Hamiltonian structures of the mCH equation to peakon weak solutions \cite{Anc-talk,ChaSzm17b}. 
But surprisingly, 
the corresponding Hamiltonian functional 
fails to reduce to a conserved integral for the $N$-peakon dynamical system \eqref{mCHpeakonsys}. 
Specifically, under reduction, 
this Hamiltonian functional yields the energy expression 
$E=\tfrac{1}{2}\sum_{i,j=1}^{N} a_i a_j \exp(-|x_i-x_j|)$
but $\dot E\neq 0$ for $2$-peakon solutions,
as noticed recently in \cite{ChaSzm16,Anc-talk}. 
This means that at least some of the integrability structure of the mCH equation,
which holds for smooth solutions, is lost for $N$-peakon weak solutions. 

More generally,  
for the mCH $N$-peakon dynamical system \eqref{mCHpeakonsys}, 
no Hamiltonian $H(a_i,x_i)$ has been found to-date to yield this system 
as Hamiltonian equations of motion 
$\dot a_i = \{ a_i,H\}$, $\dot x_i = \{ x_i,H\}$. 
The purpose of the present paper is address this open question. 

First, 
a simplified derivation of the mCH $N$-peakon dynamical system 
will be given in \secref{mCHpeakons}. 
Next, in \secref{reduction}, 
the reduction of the two Hamiltonian functionals for the mCH equation 
will be shown explicitly to yield functions $E_1(a_i,x_i)$ and $E_2(a_i,x_i)$ 
that are not conserved for this dynamical system, $\dot E_1\neq 0$ and  $\dot E_2\neq 0$,
for all $N\geq 2$.  
Then a general necessary condition is derived  
for a Poisson bracket associated to a Hamiltonian structure of a nonlinear dispersive wave equation 
to have a well-defined reduction to the dynamical variables $(a_i,x_i)$
for peakon weak solutions. 
This condition is used to provide a derivation of the non-canonical Poisson bracket \eqref{PB} 
for the mCH equation, 
using its nonlocal Hamiltonian structure. 
Moreover, the local Hamiltonian structure of the mCH equation is shown to not admit 
a reduction to the dynamical variables $(a_i,x_i)$. 

This motivates a systematic search for conserved integrals and Hamiltonians in \secref{FIsearch}. 
As the first main result, 
an explicit conserved integral is derived for all $N\geq 2$,
and its invariance properties are discussed. 
In \secref{mCHhamilstruct}, 
as the second main result, 
this conserved integral is shown to also be a Hamiltonian 
for the mCH $N$-peakon dynamical system if $N$ is even, 
by using the non-canonical Poisson bracket \eqref{PB} for peakon solutions. 
But if $N$ is odd, when this Poisson bracket is used, 
the conserved integral fails to be a Hamiltonian 
since it is shown to yield Hamiltonian equations of motion that do not agree with the dynamical system \eqref{mCHpeakonsys} for mCH $N$-peakon weak solutions. 

These results indicate that the bi-Hamiltonian structure of the mCH equation for smooth solutions 
is not preserved for weak solutions, in contrast to the situation for the CH equation. 
From a broader point of view, 
this suggests that the cubic nonlinearity of the mCH equation 
versus the quadratic nonlinearity of the CH equation 
makes a decisive change in the properties of weak solutions. 

Some concluding remarks, 
including approaches for investigation of possible Hamiltonian structures 
in the case when $N$ is odd, 
as well as the connection between the results here 
and the interesting work in \Ref{ChaSzm16,ChaSzm17a} 
on regularization conditions needed for reduction of a Lax pair to non-smooth solutions,
are discussed in \secref{concl}.

\section{mCH peakons as weak solutions}
\label{mCHpeakons} 

Weak solutions $u(x,t)$ are distributions that satisfy an integral formulation \eqref{fg-fam-weak} 
of a given peakon equation $m_t + f(u,u_x)m +(g(u,u_x)m)_x=0$ for $m=u-u_{xx}$
in some suitable function space on $\Rnum\times\Rnum$. 
For the mCH equation \eqref{mCH},
an integral formulation is obtained by, first, 
multiplying the equation by a test function $\psi(x,t)$ 
and integrating over $(x,t)\in\Rnum\times\Rnum$,
and next, using integration by parts to remove all terms involving $u_{xx}$ and $u_{xxx}$:
\begin{equation}\label{mCHweak}
\iint_{\Rnum\times\Rnum} \big( 
\psi u_t -\psi_x (u^2+u_x^2)u -\psi_{xx} (u^2u_x-\tfrac{1}{3}u_x^3 +u_t) 
\big) \, dx \, dt =0 . 
\end{equation}
All classical (smooth) solutions of the mCH equation
on the real line, $x\in\Rnum$, 
satisfy this integral equation \eqref{mCHweak}. 
Its non-smooth (distributional) solutions in the Sobolev space $W^{3,1}_\loc(\Rnum)$
are weak solutions of the mCH equation.

This weak formulation regularizes product of distributions in a certain natural way that is useful for analysis, such as proving well-posedness, local and global existence, stability, and wave breaking. 
In particular, 
local well-posedness and local existence for the Cauchy problem of the mCH equation,
as well as singularity formation and wave breaking criteria, 
have been established in \Ref{LiuOlvQuZha,GuiLiuOlvQu},
and orbital stability of peakons has been proved in \Ref{LiuLiuOlvQu}. 

$N$-peakon weak solutions \eqref{Npeakon} are 
a linear superposition of peakons $a_i(t) \exp(-|x-x_i(t)|)$, $i=1,2,\ldots, N$, 
with time-dependent amplitudes $a_i(t)$ and positions $x_i(t)$. 
A derivation of the dynamical system \eqref{mCHpeakonsys} 
for $(a_i,x_i)$ first appears in \Ref{GuiLiuOlvQu},
using an ordering assumption $x_1>x_2>\cdots>x_N$ for the positions
and evaluating the integral equation \eqref{mCHweak} by dividing up 
the integration over $x\in\Rnum$ into intervals defined by these positions. 
The derivation can be considerably simplified by employing distributional methods
which do not require any ordering or any dividing up of the integration domain,
as explained in \Ref{RecAnc}. 
This leads directly to the dynamical system \eqref{mCHpeakonsys}, 
which can also be obtained from the dynamical system \eqref{fg-peakonsys} 
for a general multi-peakon equation \eqref{fg-fam} 
by substituting $F=\int f\,d u_x=0$ and $G=\int g\,d u_x = u^2u_x -\tfrac{1}{3}u_x^3$,
where $f=0$ and $g=u^2-u_x^2$ are given by the form of the mCH equation \eqref{mCH}. 

\begin{prop}
The mCH $N$-peakon dynamical system \eqref{mCHpeakonsys} 
for weak solutions is given by 
\begin{subequations}\label{Npeakonsys}
\begin{align}
& \dot a_i = 0,
\label{Npeakon-a-eom}
\\
& \dot x_i = \tfrac{2}{3} a_i^2 + 2 a_i \sum_{j \neq i} a_j \exp(-|x_{ij}|) + 2\sum_{\substack{j <k\\j,k \neq i}} a_j a_k (1-\sgn(x_{ij}) \sgn(x_{ik}))\exp(-|x_{ij}|-|x_{ik}|),
\label{Npeakon-x-eom}
\end{align}
\end{subequations}
where 
\begin{equation}
x_{ij} = x_i - x_j
\end{equation}
with all summation indices running from $1$ to $N$. 
\end{prop}

If an ordering $x_j>x_k$ for $j<k$ is assumed, 
then this system \eqref{Npeakonsys} becomes 
\begin{equation}\label{orderedNpeakonsys}
\dot a_i = 0,
\quad
\dot x_i = \tfrac{2}{3} a_i^2 + 2 a_i \sum_{j \neq i} a_j \exp(-|x_{ij}|) + 4\sum_{j <i<k} a_j a_k \exp(-x_{jk}),
\end{equation}
since 
$|x_{jk}|=\sgn(x_{jk})x_{jk}$ 
with 
$\sgn(x_{jk})=\sgn(k-j)$
due to the ordering. 
(The ordered system given in \Ref{GuiLiuOlvQu} is missing the condition $j\neq i$ in the first sum.)
Compared to the ordered system, 
the general (unordered) system \eqref{Npeakonsys} is more useful 
for analyzing the peakon dynamics because the relative positions of peakons can interchange when a collision occurs. 

For the sequel, 
it will be useful to write out the cases $N=2$ and $N=3$ explicitly. 
The $2$-peakon dynamical system is given by 
\begin{subequations}\label{N=2peakonsys}
\begin{gather}
\dot a_1 = 0, 
\quad
\dot a_2 = 0, 
\label{N=2eoma}
\\
\dot x_1 = \tfrac{2}{3} {a_1}^2 + 2 a_1 a_2 e^{-|x_{12}|}, 
\quad
\dot x_2 = \tfrac{2}{3} {a_2}^2 + 2 a_1 a_2 e^{-|x_{12}|}, 
\label{N=2eomx}
\end{gather}
\end{subequations}
and the $3$-peakon dynamical system is given by 
\begin{subequations}\label{N=3peakonsys}
\begin{gather}
\dot a_1 = 0, 
\quad
\dot a_2 = 0, 
\quad
\dot a_3 = 0, 
\\
\dot x_1 = \tfrac{2}{3} {a_1}^2 + 2 a_1 (a_2 e^{-|x_{12}|} + a_3 e^{-|x_{31}|}) + 2 a_2 a_3 (1+s_{12}s_{31})e^{-|x_{12}|-|x_{13}|}, 
\\
\dot x_2 = \tfrac{2}{3} {a_2}^2 + 2 a_2 (a_1 e^{-|x_{12}|} + a_3 e^{-|x_{23}|}) + 2 a_1 a_3 (1+s_{12}s_{23})e^{-|x_{12}|-|x_{23}|}, 
\\
\dot x_3 = \tfrac{2}{3} {a_3}^2 + 2 a_3 (a_1 e^{-|x_{31}|} + a_2 e^{-|x_{23}|}) + 2 a_1 a_2 (1+s_{31}s_{23})e^{-|x_{31}|-|x_{23}|}, 
\end{gather}
\end{subequations}
where 
\begin{equation}
s_{ij} = \sgn(x_{ij}) . 
\end{equation}

\section{Reduction of Hamiltonian structures}
\label{reduction}

The mCH equation \eqref{mCH} has two compatible Hamiltonian structures \cite{OlvRos,Qia}
\begin{equation}
m_t = -((u^2-u_x^2)m)_x
= \H(\delta H_1/\delta m) = \E(\delta H_2/\delta m)
\end{equation}
as given by the Hamiltonian operators 
\begin{equation}\label{HEops}
\H = - \partial_x m \partial_x\inv m \partial_x, 
\qquad
\E = \partial_x^3 - \partial _x , 
\end{equation}
and the corresponding Hamiltonian functionals
\begin{equation}\label{H1H2}
H_1 = \int_{\Rnum} m u \, d x,
\qquad
H_2 = \tfrac{1}{4} \int_{\Rnum} u (u^2 - u_x^2) m \, d x . 
\end{equation}
Recall \cite{Olv-book}, 
a linear operator $\D$ is Hamiltonian if it defines an associated Poisson bracket
\begin{equation}\label{bracket}
\{F,G\}_{\D} = \int_{\Rnum} (\delta F/\delta m)\mathcal{D}(\delta G/\delta m) \, d x
\end{equation}
which is bilinear, skew, and satisfies the Jacobi identity, 
for all functionals $F,G$. 
Skew-symmetry of the Poisson bracket is equivalent to skew-symmetry $\D^* = -\D$ of the linear operator. 

In terms of the two Poisson brackets arising from the two Hamiltonian operators \eqref{HEops}, 
the mCH equation can be expressed as a bi-Hamiltonian evolution equation 
\begin{equation}\label{mCHbihamil}
m_t = \{m,H_1\}_{\H} = \{m,H_2\}_{\E} 
\end{equation}
since 
$\{m(x,t),F\}_{\D} = \int_{\Rnum} \delta(x-y)\big(\D(\delta F/\delta m)\big)\big|_{u(x,t)=u(y,t)}\, d y = \D(\delta F/\delta m)$
where $\delta(x)$ denotes the Dirac delta distribution. 
The skew-symmetry of the Hamiltonian operators is well-known to imply that 
the Hamiltonians $H_1$, $H_2$ are conserved 
for smooth solutions $u(x,t)$ of the mCH equation 
under appropriate decay conditions for $|x|\rightarrow\infty$. 

Both $H_1$ and $H_2$ have a well-defined reduction 
for all $N$-peakon solutions \eqref{Npeakon} of the mCH equation. 
Firstly, write $u=\sum_i u_i$ and $m=\sum_i m_i$ 
where 
\begin{equation}\label{ith-u-m}
u_i= a_i e^{-|x-x_i|},
\quad
u_{ix} = -a_i \sgn(x-x_i) e^{-|x-x_i|},
\quad
m_i = u_i -u_{ixx} 
= 2a_i \delta(x-x_i)
\end{equation}
as obtained through the properties 
$\sgn(x-x_i)^2=1$ and $f(u_j) \delta(x-x_i) = f(a_je^{-|x_{ji}|})\delta(x-x_i)$ 
which hold in the sense of distributions. 
Secondly, use the distribution property of $\delta(x-x_i)$ to evaluate 
\begin{equation}\label{H1peakon}
H_1 = \sum_{i,j} \int_{\Rnum} u_i m_j \, d x 
= 2 \sum_{i,j} a_i a_j \int_{\Rnum} e^{-|x-x_i|} \delta(x-x_j) \, d x 
= 2 \sum_{i,j} a_i a_j e^{-|x_{ij}|} 
:= E_1(a_i,x_i)
\end{equation}
and
\begin{equation}\label{H2peakon}
\begin{aligned}
H_2 
& = \tfrac{1}{4} \sum_{i,j,k,l} \int_{\Rnum} u_i (u_j u_k - u_{jx} u_{kx}) m_l \, d x 
\\
& = \tfrac{1}{2} \sum_{i,j,k,l} a_i a_j a_k a_l  \int_{\Rnum} (1 - \sgn(x-x_j)\sgn(x-x_k))  e^{-|x-x_i| -|x-x_j|-|x-x_k|} \delta(x-x_l) \, d x 
\\
& = \tfrac{1}{2} \sum_{i,j,k,l} a_i a_j a_k a_l  (1 - \sgn(x_{lj})\sgn(x_{lk}))  e^{-|x_{lj}| -|x_{lk}|-|x_{li}|} 
:= E_2(a_i,x_i) . 
\end{aligned}
\end{equation}
These reduced Hamiltonians $E_1(a_i,x_i)$ and $E_2(a_i,x_i)$ 
are positive-definite functions of the dynamical variables $(a_i,x_i)$. 
As will now be shown, surprisingly, 
they are not conserved for solutions of the $N$-peakon dynamical system \eqref{Npeakonsys}. 

It will be sufficient to consider the position-ordered form \eqref{orderedNpeakonsys} 
for the peakon system. 
The corresponding form of the reduced Hamiltonians \eqref{H1peakon} and \eqref{H2peakon} 
is given by 
\begin{gather}
E_1 = 2 \sum_{i} a_i^2 + 4 \sum_{i<j} a_i a_j e^{-x_{ij}}  , 
\label{H1orderedpeakon}
\\
\begin{aligned}
E_2 & = 
\tfrac{1}{2}\sum_{i} {a_i}^4 
+\sum_{i<j} a_i a_j \big( \tfrac{3}{2}({a_i}^2+{a_j}^2) + 2a_i a_j e^{-x_{ij}} \big) e^{-x_{ij}} 
\\&\qquad
+\sum_{i<j<k} a_i a_j a_k ( 4 a_j+ 3 a_i e^{-x_{ij}} + 3 a_k e^{-x_{jk}} ) e^{-x_{ik}} 
+\sum_{i<j<k< l} 4 a_i a_j a_k a_l e^{-x_{ik} -x_{jl}} .
\end{aligned}
\label{H2orderedpeakon}
\end{gather}
Their time derivative yields, by direct calculation, 
\begin{equation}\label{dotH1}
\dot E_1 = \tfrac{4}{3} \sum_{i<j} (a_j^2 - a_i^2) a_i a_j e^{-x_{ij}}  
\end{equation}
and 
\begin{equation}
\begin{aligned}
\dot E_2 = & 
\sum_{i<j} ({a_j}^2-{a_i}^2) a_i a_j ( \tfrac{3}{2}({a_j}^2+{a_i}^2) + 4a_i a_j e^{-x_{ij}} ) e^{-x_{ij}} 
+\sum_{i<j<k} 4 ({a_k}^2-{a_i}^2) a_i {a_j}^2 a_k e^{-x_{ik}} 
\\&\qquad
+\sum_{i<j<k} 3 \big( ({a_j}^2+{a_k}^2 -2{a_i}^2){a_i}^2 a_j a_k e^{-x_{ij}} 
+({a_j}^2+{a_i}^2-2{a_k}^2) a_i a_j {a_k}^2 e^{-x_{jk}} \big) e^{-x_{ik}} 
\\&\qquad
+\sum_{i<j<k< l} 4 ({a_k}^2+{a_l}^2 -{a_i}^2-{a_j}^2) a_i a_j a_k a_l e^{-x_{ik} -x_{jl}} .
\end{aligned}
\end{equation}
This establishes that $\dot E_1\neq 0$ and $\dot E_2\neq 0$ 
for $N$-peakon weak solutions that contain different positive (or negative) amplitudes, 
namely, $a_i>a_j>0$ (or $a_i<a_j<0$) for all $i>j$. 

A Poisson bracket \eqref{bracket} can be formally reduced to peakon weak solutions 
by applying a general version of a method given in \Ref{HolHon}
using the integral 
\begin{equation}\label{mmbracket}
\iint_{\Rnum^2} \{m(x,t),m(y,t)\}_{\D}\Psi(x,y)\, dx\, dy
\end{equation}
with $\Psi(x,y)$ being a test function that is skew-symmetric in $x,y$. 
The general method consists of evaluating the bracket \eqref{mmbracket} 
in the following two ways,
where 
\begin{equation}\label{Npeakon-m}
m=\sum_i m_i =2\sum_i a_i \delta(x-x_i)
\end{equation} 
is given by expression \eqref{ith-u-m}
for $N$-peakons \eqref{Npeakon}. 

Firstly, 
substitution of the $N$-peakon expression \eqref{Npeakon-m}
into a general Poisson bracket \eqref{bracket} yields 
\begin{equation}
\{m(x,t),m(y,t)\}_{\D}
= 4 \sum_{i,j} \{ a_i \delta(x-x_i), a_j \delta(y-x_j) \}_{\D} 
\end{equation}
by bilinearity of the bracket. 
The main reduction step now consists of asserting that,
for any functions $f(a_i,x_i)$ and $g(a_i,x_i)$, 
the bracket $\{ f(a_i,x_i),g(a_j,x_j) \}_{\D}$ 
represents a Poisson bracket on the space of the dynamical variables $(a_i,x_i)$, $i=1,2,\ldots,N$.
This means, in particular, that 
\begin{equation}
\begin{aligned}
\{ f,g \}_{\D} & = \sum_{i,j} \Big( 
\{ a_i, a_j \}_{\D} \Parder{f}{a_i} \Parder{g}{a_j} 
+ \{ a_i, x_j \}_{\D} \Parder{f}{a_i} \Parder{g}{x_j} 
\\&\qquad\qquad
+\{ x_i, a_j \}_{\D} \Parder{f}{x_i} \Parder{g}{a_j} 
+ \{ x_i, x_j \}_{\D} \Parder{f}{x_i} \Parder{g}{x_j} 
\Big)
\end{aligned}
\end{equation}
can be assumed to hold where 
$\{ a_i, a_j \}_{\D}$, $\{ a_i, x_j \}_{\D}$, $\{ x_i, a_j \}_{\D}$, $\{ x_i, x_j \}_{\D}$ 
are regarded as defining the structure functions of a Poisson bracket. 
Hence, 
\begin{equation}
\begin{aligned}
& \{ a_i \delta(x-x_i), a_j \delta(y-x_j) \}_{\D} 
\\
& = \{ a_i, a_j \}_{\D} \delta(x-x_i) \delta(y-x_j) 
- \{ a_i, x_j \}_{\D} a_j\delta(x-x_i) \delta'(y-x_j) 
\\&\qquad
- \{ x_i, a_j \}_{\D} a_i\delta'(x-x_i) \delta(y-x_j) 
+ \{ x_i, x_j \}_{\D} a_i a_j\delta'(x-x_i) \delta'(y-x_j)
\end{aligned}
\end{equation}
which yields
\begin{equation}
\begin{aligned}
& \{m(x,t),m(y,t)\}_{\D}
\\
&= 4 \sum_{i,j} \big( 
 \{ a_i, a_j \}_{\D} \delta(x-x_i) \delta(y-x_j) 
- a_j\{ a_i, x_j \}_{\D} \delta(x-x_i) \delta'(y-x_j) 
\\&\qquad
- a_i\{ x_i, a_j \}_{\D} \delta'(x-x_i) \delta(y-x_j) 
+ a_ia_j\{ x_i, x_j \}_{\D} \delta'(x-x_i) \delta'(y-x_j)
\big) .
\end{aligned}
\end{equation}
Then the integral \eqref{mmbracket} becomes 
\begin{equation}\label{PBreduction}
\begin{aligned}
\iint_{\Rnum^2} \{m(x,t),m(y,t)\}_{\D}\Psi(x,y)\, dx\, dy
= 4 \sum_{i\neq j} & \big( 
\{ a_i, a_j \}_{\D} \Psi(x_i,x_j) 
+ a_ia_j\{ x_i, x_j \}_{\D} \Psi_{xy}(x_i,x_j) 
\\&\qquad
+ a_j(\{ a_i, x_j \}_{\D} -\{ x_j, a_i \}_{\D})\Psi_y(x_i,x_j) 
\big)
\end{aligned}
\end{equation}
after integration by parts and use of the properties 
$\Psi(y,x)=-\Psi(x,y)$, $\Psi_y(y,x)=-\Psi_x(x,y)$, and $\Psi_{xy}(y,x)=-\Psi_{xy}(x,y)$
which hold due to the skew-symmetry of $\Psi$. 
This reduction of the bracket holds for any Hamiltonian operator $\D$. 

Secondly, 
the explicit formula \eqref{bracket} for a Poisson bracket defined 
in terms of a Hamiltonian operator $\D$ yields 
\begin{equation}
\{m(x,t),m(y,t)\}_{\D}
= \int_{\Rnum} \delta(x-z) \D( \delta(y-z) )\, dz = \D \delta(y-x) . 
\end{equation}
Then the integral \eqref{mmbracket} is given by 
\begin{equation}\label{PBeval}
\begin{aligned}
\iint_{\Rnum^2} \{m(x,t),m(y,t)\}_{\D}\Psi(x,y)\, dx\, dy
& = \iint_{\Rnum^2} \Psi(x,y) \D(\delta(y-x)) \, dx\, dy
\\
& = \int_{\Rnum} \big({-\D}\Psi(x,y) \big)\big|_{y=x}\, dx
\end{aligned}
\end{equation}
after the skew-symmetry of $\D$ is used. 

This establishes the following useful general reduction result.

\begin{lem}\label{PB-reduction-structure}
If the Poisson bracket \eqref{bracket} associated to a Hamiltonian operator $\D$ 
has a well-defined reduction to the space of the dynamical variables 
$(a_i,x_i)$, $i=1,2,\ldots,N$, 
arising from $N$-peakon weak solutions \eqref{Npeakon} for all $N\geq 1$, 
then the structure functions 
$\{ a_i, a_j \}_{\D}$, $\{ a_i, x_j \}_{\D}$, $\{ x_i, a_j \}_{\D}$, $\{ x_i, x_j \}_{\D}$ 
of the reduced Poisson bracket must satisfy the equation
\begin{equation}\label{PBreduction-condition}
\begin{aligned}
4 \sum_{i\neq j} & \big( 
\{ a_i, a_j \}_{\D} \Psi(x_i,x_j) 
+ a_ia_j\{ x_i, x_j \}_{\D} \Psi_{xy}(x_i,x_j) 
+ a_j(\{ a_i, x_j \}_{\D} -\{ x_j, a_i \}_{\D})\Psi_y(x_i,x_j) 
\big)
\\& = \int_{\Rnum} \big({-\D}\Psi(x,y) \big)\big|_{y=x}\, dx
\end{aligned}
\end{equation}
for an arbitrary test function $\Psi(x,y)$ that is skew-symmetric in $x,y$. 
\end{lem}

Now, equation \eqref{PBreduction-condition} can be used to determine 
the structure functions when they exist. 
However, since the one side \eqref{PBreduction-condition} contains 
$\Psi$ and its derivatives evaluated at pairs of distinct points $(x_i,x_j)$, $i\neq j$, 
the other side \eqref{PBeval} needs to have the same property. 
As a consequence, 
when $u$, $u_x$, $m$ are given by the peakon expressions \eqref{ith-u-m}, 
the coefficients in the linear operator $\D$ must reduce to distributions 
whose support consists only of the points $x_i$, $i=1,2,\ldots,N$. 

For the pair of mCH Hamiltonian operators \eqref{HEops}, 
this distributional support condition is satisfied by the first Hamiltonian operator $\H$, 
but not by the second Hamiltonian operator $\E$. 
Therefore, 
the second mCH Poisson bracket cannot be reduced to peakon weak solutions. 

The reduction of the first mCH Poisson bracket is obtained by evaluating the integral 
\begin{equation}
\int_{\Rnum} \big({-\H}\Psi(x,y) \big)\big|_{y=x}\, dx
= \int_{\Rnum} \big( \partial_x(m\partial_x\inv (\Psi_x(x,y)m)) \big)\big|_{y=x}\, dx . 
\end{equation}
Substitution of $m=\sum_{i} m_i$ given by the expression \eqref{ith-u-m}, 
followed by use of the relation $\partial\inv_x (\delta(x-z) f(x)) = \theta(x-z)f(x)$
where $\theta(z)$ denotes the Heaviside step function, 
yields 
\begin{equation}
\begin{aligned}
\int_{\Rnum} \big( \partial_x(m\partial_x\inv (m \Psi_x(x,y))) \big)\big|_{y=x}\, dx 
& = 4 \sum_{i,j} a_i a_j \int_{\Rnum} \partial_x\big( \delta(x-x_i) \theta(x-x_j) \big) \Psi_x(x_j,x)\, dx 
\\
& = {-4} \sum_{i,j} a_i a_j \theta(x_{ij}) \Psi_{xy}(x_j,x_i) 
\end{aligned}
\end{equation}
after integration by parts. 
This result can be simplified by use of 
the skew-symmetry property $\Psi_{xy}(y,x)=-\Psi_{xy}(x,y)$
combined with the identity $\theta(z)=\tfrac{1}{2}(1+\sgn(z))$,
which gives
$\sum_{i,j} a_i a_j \theta(x_{ij}) \Psi_{xy}(x_j,x_i) = -\tfrac{1}{2}\sum_{i,j} a_i a_j \sgn(x_{ij}) \Psi_{xy}(x_i,x_j) $,
and hence
\begin{equation}\label{Hopbracket}
\int_{\Rnum} \big({-\H}\Psi(x,y) \big)\big|_{y=x}\, dx
= 2\sum_{i,j} a_i a_j \sgn(x_{ij}) \Psi_{xy}(x_i,x_j) .
\end{equation}
When this integral \eqref{Hopbracket} is substituted into the reduction equation \eqref{PBreduction-condition},
and the separate coefficients of $\Psi(x_i,x_j)$, $\Psi_y(x_i,x_j)$, $\Psi_{xy}(x_i,x_j)$ 
are equated on both sides, 
the resulting structure functions yield the Poisson bracket structure \eqref{PB}
defined for peakon solutions of the mCH equation. 

The same situation occurs for the CH equation: 
just one of its two Hamiltonian operators satisfies the distributional support condition 
needed for the associated Poisson bracket to have a well-defined reduction 
via Lemma~\ref{PB-reduction-structure}.

\section{A conserved integral for mCH $N$-peakon weak solutions}
\label{FIsearch}

It is simplest to begin by considering 
the mCH $N$-peakon dynamical system \eqref{Npeakonsys} for $(a_i,x_i)$ 
in the case $N=2$. 
This system consists of a pair of 
coupled nonlinear ODEs \eqref{N=2eomx}  for $x_1$ and $x_2$, 
with $a_1$ and $a_2$ being arbitrary constants. 
The most important feature of these equations is that 
they are invariant under position shifts $x_1\to x_1 +\epsilon$ and $x_2\to x_2 +\epsilon$
where $\epsilon$ is an arbitrary constant. 
This invariance implies that the variable $x_{12}=x_1-x_2$ 
satisfies a decoupled ODE $\dot x_{12} = \tfrac{2}{3} ({a_1}^2 - {a_2}^2)$,
which yields 
\begin{equation}\label{2peakonsol-x12}
x_{12} = \tfrac{2}{3} ({a_1}^2 - {a_2}^2) t + C, 
\quad
C=\const . 
\end{equation}
Then the pair of ODEs for $x_1$ and $x_2$ can be directly integrated,
obtaining 
\begin{equation}\label{2peakonsol-x1-x2}
\begin{aligned}
& 
x_1 = \tfrac{2}{3} {a_1}^2 t - X(t) +\tfrac{1}{2}C , 
\quad
x_2 = \tfrac{2}{3} {a_2}^2 t - X(t) -\tfrac{1}{2}C , 
\\
& 
X(t) = \frac{3a_1 a_2}{a_1^2 - a_2^2}\sgn(x_{12}) e^{-|x_{12}|} +\tilde C, 
\quad
\tilde C=\const 
\end{aligned}
\end{equation}
in the general case when $|a_1|\neq |a_2|$. 
The two arbitrary constants $C$ and $\tilde C$ 
in this solution \eqref{2peakonsol-x12}--\eqref{2peakonsol-x1-x2}
can be expressed in terms of $t,x_1,x_2$: 
\begin{gather}
C=I_1 = \tfrac{2}{3} ({a_2}^2 - {a_1}^2) t + x_1-x_2, 
\\
\tilde C = I_2 = \tfrac{1}{3}({a_1}^2 + {a_2}^2)t  -\tfrac{1}{2}(x_1+x_2) 
-\frac{3a_1 a_2}{{a_1}^2 - {a_2}^2}\sgn(x_{12}) e^{-|x_{12}|} .
\end{gather}
These expressions represent conserved integrals, 
which satisfy $\dot I_1=\dot I_2=0$ for all $t$ such that $x_{12}\neq 0$. 
Their linear combination 
\begin{equation}
I= ({a_2}^2-{a_1}^2)I_2 -\tfrac{1}{2}({a_1}^2+{a_2}^2)I_1 
= {a_1}^2 x_2 -{a_2}^2 x_1 +3a_1 a_2\sgn(x_{12}) e^{-|x_{12}|}
\end{equation}
is a local conserved integral having no explicit dependence on $t$. 
Globally, $I$ is continuous in $t$ except at the time 
$t=\tfrac{3}{2}I_1/({a_2}^2 -{a_1}^2)$ corresponding to $x_{12}=0$, 
namely, at the time of a collision between the two peakons,
where $I$ has a jump discontinuity. 

Thus, $I$ is a local constant of motion for the solution  \eqref{2peakonsol-x12}--\eqref{2peakonsol-x1-x2}. 
More generally, 
\begin{equation}
\dot I = {a_1}^2 \dot x_2 -{a_2}^2 \dot x_1 -3a_1 a_2e^{-|x_{12}|} \dot x_{12} =0
\end{equation}
holds locally in $t$ for all solutions of the ODEs \eqref{N=2eomx} for $x_1$ and $x_2$,
including the special case when $|a_1|=|a_2|$. 

Therefore, $I$ is a local constant of motion for all $2$-peakon solutions. 
It has the feature that its parity under permutation of $(a_1,x_1)$ and $(a_2,x_2)$ is odd. 
If $I$ is multiplied by $\sgn(x_{12})$,
which has odd parity and which is locally constant with respect to $t$, 
then this yields 
\begin{equation}\label{N=2peakonFI}
I_{(2)}= \sgn(x_{12})({a_1}^2 x_2 -{a_2}^2 x_1) +3a_1 a_2 e^{-|x_{12}|},
\quad
\dot I_{(2)}=0 
\end{equation}
which is a permutation-invariant local constant of motion for all $2$-peakon solutions. 
An interesting observation is that $I_{(2)}$ is not invariant under position shifts. 

Since the mCH $2$-peakon dynamical system \eqref{N=2peakonsys} 
has four dynamical variables, 
it admits three functionally-independent local constants of motion, 
and one local conserved integral that explicitly contains $t$,
all of which can be chosen to be permutation-invariant. 
In particular, 
the constants of motion $I_{(2)}$, $a_1+a_2$, $a_1a_2$ 
are functionally independent and permutation invariant. 
This establishes the following result. 

\begin{prop}
Every permutation-invariant local constant of motion of the mCH $2$-peakon dynamical system \eqref{N=2peakonsys} 
is given by a function $I=I(a_1+a_2,a_1a_2,I_{(2)})$.
\end{prop}
 
By comparison, 
note that the reduced mCH Hamiltonians \eqref{H1peakon} and \eqref{H2peakon}
are given by 
\begin{equation}
E_1 = {a_1}^2 + {a_2}^2 + 2a_1 a_2 e^{-|x_{12}|},
\quad
E_2 = a_1 a_2 (a_1 +a_2)^2 e^{-|x_{12}|}
\end{equation}
in the case $N=2$. 
Both of these functions are invariant under positions shifts, 
and hence they cannot be expressed in the form $I(a_1+a_2,a_1a_2,I_{(2)})$.
This simple observation directly proves that no function of $E_1$ and $E_2$
produces a conserved integral for the mCH $2$-peakon dynamical system \eqref{N=2peakonsys}. 

The form of the conserved integral \eqref{N=2peakonFI} for $2$-peakon solutions 
suggests a search for a general conserved integral 
$I=\sum_{i\neq j} \big( \alpha_{ij} {a_j}^2 x_i + \beta_{ij} a_i a_j e^{-|x_{ij}|} \big)$ 
for $N$-peakon weak solutions for all $N\geq 2$, 
where the coefficients $\alpha_{ij}$, $\beta_{ij}$ are constants to be determined. 
This search yields the following result. 

\begin{thm}\label{conserved}
For all $N\geq 2$, 
the function 
\begin{equation}\label{NpeakonFI}
I_{(N)} = \sum_{i\neq j} \big( \sgn(x_{ij}) (-1)^{s_j + s_i} a_j^2 x_i + \tfrac{3}{2} a_i a_j e^{-|x_{ij}|} \big) 
\end{equation}
is a permutation-invariant local constant of motion of the mCH $N$-peakon dynamical system \eqref{Npeakonsys}. 
Here $s_k$ denotes the number of signs that are equal to $+1$ in the set 
$(\sgn(x_{1k}),\ldots$, $\sgn(x_{Nk}))$. 
Moreover, for all odd $N\geq3$, 
$I_{(N)}$ is also invariant under position shifts, 
$x_i\to x_i+\epsilon$, $i=1,\ldots,N$. 
When the system is expressed in the position-ordered form \eqref{orderedNpeakonsys}, 
this constant of motion is given by 
\begin{equation}\label{NpeakonFIordered}
I_{(N)} = \begin{cases}
\sum_{i<j} \big( 3 a_i a_j e^{-x_{ij}} + (-1)^{i+j} (a_j^2 x_i - a_i^2 x_j) \big), 
& N\geq 2
\\
\sum_{i<j} 3 a_i a_j e^{-x_{ij}} - \sum_{i\text{ even}}\big( x_{i-1\, i}\sum_{j\geq i} (-1)^{j} a_j^2  + x_{i\, i+1}\sum_{j\leq i} (-1)^{j} a_j^2 \big), 
& N\text{ odd} . 
\end{cases}
\end{equation}
\end{thm}

The proof is a straightforward albeit somewhat lengthy computation,
which is given in the Appendix.

\section{A Hamiltonian structure}
\label{mCHhamilstruct}

The permutation-invariant local constant of motion \eqref{NpeakonFI} 
is an obvious candidate for a Hamiltonian $H=\lambda I_{(N)}$ 
of the mCH $N$-peakon dynamical system \eqref{Npeakonsys},
where $\lambda$ is a normalization constant. 
From the non-canonical Poisson bracket structure \eqref{PB}, 
the Hamiltonian equations of motion 
are given by 
\begin{align}
\dot a_i = \lambda\{ a_i, I_{(N)} \} 
& = \lambda\sum_{j} \Big( \{ a_i, a_j\} \Parder{I_{(N)}}{a_j} + \{ a_i, x_j\} \Parder{I_{(N)}}{x_j} \Big) 
=0, 
\label{NpeakFI-eoma}
\\
\dot x_i = \lambda\{ x_i, I_{(N)} \} 
& = \lambda\sum_{j} \Big( \{ x_i, a_j\} \Parder{I_{(N)}}{a_j} + \{ x_i, x_j\} \Parder{I_{(N)}}{x_j} \Big) 
\nonumber\\
& = \tfrac{1}{2}\lambda \sum_{k\neq j} \sgn(x_{ij}) \sgn(x_{jk}) \big( (-1)^{s_j + s_k} {a_k}^2 - 3 a_k a_j e^{-|x_{jk}|} \big) .
\label{NpeakFI-eomx}
\end{align}

When $N=2$, these equations of motion for $x_i$ become
\begin{equation}
\dot x_1 = \tfrac{\lambda}{2} ( {a_1}^2 + 3 a_1 a_2 e^{-|x_{12}|} ), 
\quad
\dot x_2 = \tfrac{\lambda}{2} ( {a_2}^2 + 3 a_1 a_2 e^{-|x_{12}|} ), 
\label{N=2FIeomx}
\end{equation}
which are seen to coincide with the equations 
in the $N=2$ peakon dynamical system \eqref{N=2peakonsys} 
if $\lambda = \tfrac{4}{3}$. 

But when $N=3$, 
this agreement no longer holds. 
In particular, 
if an ordering $x_j>x_k$ for $j<k$ is assumed, 
then the equations of motion \eqref{NpeakFI-eomx} are given by 
\begin{subequations}\label{N=3FIeomx}
\begin{align}
& \dot x_1 = \tfrac{\lambda}{2} ( {a_2}^2 -{a_3}^2 + 3 a_1 (a_2 e^{-x_{12}} + a_3 e^{-x_{13}}) ), 
\\
& \dot x_2 = \tfrac{\lambda}{2} ( 2{a_2}^2 -{a_1}^2 -{a_3}^2 + 3 a_2 (a_1 e^{-x_{12}} + a_3 e^{-x_{23}}) +6a_1a_3 e^{-x_{13}} ),
\\
& \dot x_3 = \tfrac{\lambda}{2} ( {a_2}^2 -{a_1}^2 + 3 a_3 (a_1 e^{-x_{13}} + a_2 e^{-x_{23}}) ) . 
\end{align}
\end{subequations}
These equations each differ from the corresponding equations 
in the $N=3$ ordered peakon dynamical system \eqref{N=3peakonsys} 
by the term $\tfrac{\lambda}{2}({a_1}^2-{a_2}^2+{a_3}^2)$
which is non-vanishing whenever the amplitudes satisfy ${a_1}^2+{a_3}^2\neq {a_2}^2$.
 
A similar result can be shown to hold for all $N\geq 2$.

\begin{thm}\label{Hamilstruct-Neven}
The permutation-invariant local constant of motion \eqref{NpeakonFI} yields a 
Hamiltonian 
\begin{equation}\label{evenN-H}
H(a_i,x_i)= \tfrac{4}{3} I_{(N)}
\end{equation}
for the mCH $N$-peakon dynamical system \eqref{Npeakonsys}
using the Poisson bracket \eqref{PB} iff $N$ is even. 
\end{thm}

The proof amounts to a direct computation of the bracket \eqref{NpeakFI-eomx},
using the position ordering $x_1>x_2>\cdots>x_N$ without loss of generality. 
This is carried out in the Appendix.

\section{Concluding remarks}
\label{concl}

A Hamiltonian structure 
has been derived for the dynamical system \eqref{Npeakonsys} for 
mCH $N$-peakon weak solutions when $N$ is even, 
with the use of the non-canonical Poisson bracket \eqref{PB}. 
The peakon Hamiltonian \eqref{evenN-H} is a conserved integral for all $N$ 
but the Hamiltonian equations of motion arising from it do not coincide 
with the mCH $N$-peakon dynamical system \eqref{Npeakonsys} when $N$ is odd. 

Moreover, 
this Hamiltonian \eqref{evenN-H} is not given by the reduction of either of the two Hamiltonians 
coming from the bi-Hamiltonian structure of the mCH equation,
and it does not appear to arise from any of the known conservation laws of the mCH equation. 
In fact, the two mCH Hamiltonians \eqref{H1H2} fail to be conserved for $N$-peakon weak solutions when $N\geq2$. 
Since the first Hamiltonian is equivalent to the $H^1$ norm of $u(x,t)$, 
this means that the $H^1$ norm of these solutions can become unbounded (either positve or negative). 

These surprising results suggest several further directions of work. 

Is the mCH $N$-peakon dynamical system 
integrable in the Liouville sense when $N$ is even? 
Does the mCH $N$-peakon dynamical system have any Hamiltonian structure when $N$ is odd? 
More generally, 
under what conditions does a Hamiltonian structure 
for a nonlinear dispersive wave equation \eqref{fg-fam} 
reduce to a Hamiltonian for the dynamical system \eqref{fg-peakonsys}
describing $N$-peakon weak solutions? 

Clearly, the loss of conservation of the mCH Hamiltonians 
when they are reduced to $N$-peakon weak solutions 
indicates that the integrability structure of the mCH equation for smooth solutions
is not preserved for peakon weak solutions. 
This important point has been addressed recently in work \cite{ChaSzm16,ChaSzm17a,ChaSzm17b} by Chang and Szmigielski. 
They have shown that the Lax pair of the mCH equation for smooth solutions 
must be regularized in a certain way for it to be preserved for non-smooth solutions
given by distributions. 
Specifically, when $N$-peakons \eqref{Npeakon} are considered, 
$u$ is a $C^0$ function 
whereas $u_x$ is a Heaviside (step function) distribution 
and $m=u-u_{xx}$ is Dirac delta distribution,
with singularities at the positions $x_i(t)$ of the $N$ peakons. 
Since the mCH equation contains cubic nonlinearities $(u^2-u_x^2)m$, 
these terms do not make sense for $N$-peakons $u$ 
unless products of step-function and Dirac-delta distributions are regularized. 

The weak formulation \eqref{mCHweak} provides one natural regularization,
which is the standard setting for doing analysis. 
However, 
this regularization does not agree with the regularization introduced in \Ref{ChaSzm16,ChaSzm17a},
since the resulting equations of motion for $x_i(t)$ turn out to differ. 
In particular, 
the terms $\tfrac{2}{3} a_i^2$ that occur in the $N$-peakon dynamical system \eqref{Npeakonsys} 
derived using the weak formulation 
are \underline{not} present in the $N$-peakon dynamical system derived using the regularization that preserves the Lax pair. 
When these terms are absent, 
the reduced mCH Hamiltonians are actually conserved for the modified $N$ peakon dynamical system, 
and for this reason the resulting distributional solutions have been called ``conservative peakons''
in \Ref{ChaSzm16,ChaSzm17a,ChaSzm17b}. 
Moreover, due to the Lax pair being preserved, 
the conservative $N$-peakon dynamical system is Liouville integrable \cite{ChaSzm17b}.
One unusual feature of this system is that $2$-peakon solutions have trivial dynamics, 
$\dot x_1=\dot x_2 =\const$, 
namely $u(x,t)$ is a superposition of two peakon travelling waves 
with constant amplitudes and equal speeds. 
The general behaviour of conservative $N$-peakons is worked out and discussed in \Ref{ChaSzm15}. 

We remark that the standard $N$-peakon weak solutions studied in the present work 
are referred to as ``dissipative peakons'' in \Ref{ChaSzm17b}.
This is somewhat of a misnomer, 
since the $H^1$ norm of these solutions does not necessarily decrease, 
as shown by expression \eqref{dotH1}. 
In fact, the $H^1$ norm will be an increasing function of $t$ 
if the amplitudes of the $N$ peakons decrease from left to right and have the same sign. 
Moreover, the $2$-peakon solutions have non-trivial dynamics, 
such that a collision occurs whenever the two peakons have different amplitudes, $|a_1|\neq |a_2|$. 

Further discussion of the difference between peakon weak solutions and conservative peakon solutions will be provided elsewhere \cite{Anc17}.

\section*{Acknowledgements}
The authors thank Jacek Szmigielski and Xiange Chang for stimulating discussions 
on this work. 
D.K.\ thanks the Department of Mathematics and Statistics at Brock University 
for support during the period August 2016 to June 2017 when this work was completed. 
S.C.A.\ is supported by an NSERC research grant.

\section*{Appendix}

\subsection{Proof of Theorem~\ref{conserved}}

To show that $I_{(N)}$ is invariant under permutations of the set of dynamical variables $(a_i,x_i)$, 
let $x_{i_1}>\cdots >x_{i_N}$ be the ordering of the positions at a fixed time $t$,
where $(i_1,\ldots,i_N)$ denotes a permutation of $(1,\ldots,N)$. 
Note $s_{i_n}= n-1$ holds for all $n$, 
and $\sgn(x_{i_n i_m})=\sgn(m-n)$ holds for all $n<m$. 
Then the terms $\sum_{i\neq j} \sgn(x_{ij}) (-1)^{s_j + s_i} a_j^2 x_i$
can be rearranged into the form 
\begin{equation}
\begin{aligned}
\sum_{n\neq m} \sgn(x_{i_n i_m}) (-1)^{s_{i_m} + s_{i_n}} a_{i_m}^2 x_{i_n}
& = \sum_{n < m} (-1)^{m + n} a_{i_m}^2 x_{i_n} - \sum_{n > m} (-1)^{m + n} a_{i_m}^2 x_{i_n}
\\& = \sum_{n < m} (-1)^{m + n}( a_{i_m}^2 x_{i_n} - a_{i_n}^2 x_{i_m} ) .
\end{aligned}
\end{equation}
Likewise, 
the other terms $\tfrac{3}{2} \sum_{i\neq j} a_i a_j e^{-|x_{ij}|}$ 
can be rearranged into the form 
\begin{equation}
\tfrac{3}{2} \sum_{n\neq m} a_{i_n} a_{i_m} e^{-|x_{i_n i_m}|} 
= 3 \sum_{n< m} a_{i_n} a_{i_m} e^{-|x_{i_n i_m}|} .
\end{equation}
The resulting expression 
$I_{(N)}= \sum_{n < m}\big( (-1)^{m + n}( a_{i_m}^2 x_{i_n} - a_{i_n}^2 x_{i_m} ) +3 a_{i_n} a_{i_m} e^{-|x_{i_n i_m}|} \big)$ 
depends only on the relative order of the positions $x_{i_n}$
and not on how they are labeled. 
Since $I_{(N)}$ is thereby invariant under permutations of the labels $(i_1,\ldots,i_N)$, 
it is correspondingly invariant under permutations of the set of dynamical variables $(a_i,x_i)$.

Conservation $\dot I_{(N)}=0$ can be established by a direct calculation 
using the position-ordered form of the dynamical system \eqref{Npeakonsys}
as follows. 
The time derivative of the position-ordered expression \eqref{NpeakonFIordered} 
for $I_{(N)}$ is given by 
\begin{equation}
\dot I_{(N)} = \sum_{i<j} \big( ( (-1)^{i+j} {a_j}^2 - 3 a_ia_j e^{-x_{ij}} )\dot x_i 
-( (-1)^{i+j} {a_i}^2 - 3a_i a_j e^{-x_{ij}} )\dot x_j \big) . 
\end{equation}
When the position-ordered equation of motion \eqref{orderedNpeakonsys} is substituted for $\dot x_k$, 
three types of terms arise: 
ordered double sums and ordered triple sums
\begin{equation}\label{I-2sum-3sum}
 \sum_{i<j} a_i a_j({a_j}^2- {a_i}^2)  e^{-x_{ij}},
\quad
 \sum_{i<j<k} \big( (-1)^{k+j} a_i a_j {a_k}^2 e^{-x_{ij}} - (-1)^{i+j} a_j a_k {a_i}^2 e^{-x_{jk}} \big);
\end{equation}
partially ordered triple sums 
\begin{equation}\label{I-p3sum}
 \sum_{i,j<k} \big( 3 a_i a_j {a_k}^2 e^{-x_{ik}}  -(-1)^{i+k} a_k a_j {a_i}^2 \big) e^{-x_{jk}} ,
\quad
-\!\sum_{i<j,k} \big( 3 a_j a_k {a_i}^2 e^{-x_{ij}}  -(-1)^{i+j} a_k a_i {a_j}^2 \big) e^{-x_{ik}} ;
\end{equation}
partially ordered quadruple sums
\begin{equation}\label{I-p4sum}
\sum_{l,i<j<k} \big( 6 a_l a_i a_j a_k e^{-x_{ij}} -2 (-1)^{i+j} a_l a_k {a_i}^2 \big) e^{-x_{lk}} ,
\ 
-\!\!\!\sum_{l<i<j,k} \big( 6 a_l a_i a_j a_k e^{-x_{ij}} -2(-1)^{i+j} a_l a_k {a_j}^2 \big) e^{-x_{lk}} .
\end{equation}
First, 
each of these quadruple sums can be decomposed into an ordered quadruple sum, 
plus ordered triple sums. 
After cancellations of terms, 
this yields
\begin{gather}
\begin{aligned}
& \sum_{l<i<j,k} 2(-1)^{i+j} a_l a_k {a_j}^2 e^{-x_{lk}} 
-\!\!\! \sum_{l,i<j<k} 2 (-1)^{i+j} a_l a_k {a_i}^2 e^{-x_{lk}}
\\&
= 2\sum_{l<i<j<k} \big( (-1)^{i+j} a_l a_k ({a_j}^2 -{a_i}^2) e^{-x_{lk}} 
+(-1)^{i+k} a_l a_j {a_k}^2 e^{-x_{lj}} -(-1)^{l+j} a_i a_k {a_l}^2 e^{-x_{ik}} \big)
\\&\qquad
+2\sum_{i<j<k} \big( (-1)^{j+k} a_i {a_k}^3 -(-1)^{j+i} a_k {a_i}^3 \big) e^{-x_{ik}} , 
\end{aligned}
\label{I-p4sum1}
\\
\sum_{l,i<j<k} 6 a_l a_i a_j a_k e^{-x_{ij}-x_{lk}}  
-\!\!\!\sum_{l<i<j,k} 6 a_l a_i a_j a_k e^{-x_{ij}-x_{lk}} 
= 6\sum_{i<j<k} ( a_j a_k {a_i}^2 e^{-x_{ij}} -a_i a_j {a_k}^2 e^{-x_{jk}} ) e^{-x_{ik}} . 
\label{I-p4sum2}
\end{gather}
In a similar way, 
the triple sums \eqref{I-p3sum} yield 
\begin{equation}
 \sum_{i,j<k} 3 a_i a_j {a_k}^2 e^{-x_{ik}-x_{jk}} 
-\!\sum_{i<j,k} 3 a_j a_k {a_i}^2 e^{-x_{ij}-x_{ik}} 
=  6\sum_{i<j<k} \big( a_i a_j {a_k}^2 e^{-x_{jk}} - a_j a_k {a_i}^2 e^{-x_{ij}} \big) e^{-x_{ik}} 
\end{equation}
which cancels the triple sum \eqref{I-p4sum2},
and
\begin{equation}\label{I-p3sum2}
\begin{aligned}
&\sum_{i<j,k} (-1)^{i+j} a_k a_i {a_j}^2 e^{-x_{ik}} 
-\! \sum_{i,j<k} (-1)^{i+k} a_k a_j {a_i}^2 e^{-x_{jk}} 
\\&
= \sum_{i<j<k} (-1)^{i+k} ( a_i a_j {a_k}^2 e^{-x_{ij}} - a_j a_k {a_i}^2 e^{-x_{jk}} )
+ \sum_{i<j<k} ((-1)^{i+j} -(-1)^{j+k}) a_i a_k {a_j}^2 e^{-x_{ik}} 
\\&\qquad
+\sum_{i<j} (-1)^{i+j} a_i a_j( {a_j}^2 - {a_i}^2 )e^{-x_{ij}} .
\end{aligned}
\end{equation}
Next, 
the double sum in expression \eqref{I-p3sum2} can be decomposed into the parts
\begin{equation}\label{I-2sum2}
\sum_{\substack{i<j\\ i+j \text{ even}}} a_i a_j( {a_j}^2 - {a_i}^2 )e^{-x_{ij}} 
-\!\sum_{\substack{i<j\\ i+j \text{ odd}}} a_i a_j( {a_j}^2 - {a_i}^2 )e^{-x_{ij}} , 
\end{equation}
while the triple sum in expression \eqref{I-p4sum1} can be simplified to get 
a similar double sum 
\begin{equation}\label{I-2sum1}
2\sum_{i<k} \sum_{1\leq j\leq k-i-1} (-1)^j ( a_i {a_k}^3 - a_k {a_i}^3 ) e^{-x_{ik}} 
= -2\!\!\sum_{\substack{i<k\\ i+k \text{ even}}} a_i a_k( {a_k}^2 - {a_i}^2 ) e^{-x_{ik}} . 
\end{equation}
These double sums \eqref{I-2sum1} and \eqref{I-2sum2}
then combine and cancel the double sum \eqref{I-2sum-3sum}. 
A similar simplification of the quadruple sum in expression \eqref{I-p4sum1} gives 
the triple sums
\begin{equation}\label{I-3sum1}
2\!\!\!\sum_{\substack{i<j<k\\ j+k\text{ even}}} a_i a_k {a_j}^2 e^{-x_{ik}} +(-1)^{i+j} a_j a_k {a_i}^2 e^{-x_{jk}} 
-2\!\!\!\sum_{\substack{i<j<k\\ i+j\text{ even}}} a_i a_k {a_j}^2 e^{-x_{ik}} +(-1)^{i+k} a_i a_j {a_k}^2 e^{-x_{ij}} . 
\end{equation}
Next, the second triple sum in expression \eqref{I-p3sum2} can be simplified to get
\begin{equation}\label{I-3sum2}
\sum_{i<j<k} (-1)^{j}((-1)^i -(-1)^k) a_i a_k {a_j}^2 e^{-x_{ik}} 
= -2\!\!\!\sum_{\substack{i<j<k\\ i+k\text{ odd}}} (-1)^{j+k} a_i a_k {a_j}^2 e^{-x_{ik}} . 
\end{equation}
Likewise, the first triple sum in expression \eqref{I-p3sum2} 
combined with the remaining triple sum \eqref{I-2sum-3sum} 
can be simplified to get 
\begin{equation}
\begin{aligned}
& \sum_{i<j<k} ((-1)^{i+k} +(-1)^{j+k}) a_i a_j {a_k}^2 e^{-x_{ij}} 
-\sum_{i<j<k} ((-1)^{i+k} +(-1)^{i+j}) a_j a_k {a_i}^2 e^{-x_{jk}} 
\\
& = 2\!\!\!\sum_{\substack{i<j<k\\ i+j\text{ even}}} (-1)^{i+k} a_i a_j {a_k}^2 e^{-x_{ij}}
-2\!\!\!\sum_{\substack{i<j<k\\ j+k\text{ even}}} (-1)^{i+j} a_j a_k {a_i}^2 e^{-x_{jk}} ,
\end{aligned}
\end{equation}
which cancels two of the triple sums in expression \eqref{I-3sum1}. 
The other two triple sums in expression \eqref{I-3sum1}
can be decomposed into parts in which $j+k$ and $i+j$ are even and odd, 
yielding 
\begin{equation}\label{I-3sum3}
2\!\!\!\sum_{\substack{i<j<k\\ j+k\text{ even}\\i+j\text{ odd}}} a_i a_k {a_j}^2 e^{-x_{ik}}
-2\!\!\!\sum_{\substack{i<j<k\\ i+j\text{ even}\\j+k\text{ odd}}} a_i a_k {a_j}^2 e^{-x_{ik}} . 
\end{equation}
Finally, 
the triple sum \eqref {I-3sum2} can be decomposed into parts in which $j+k$ is odd and even,
which gives 
\begin{equation}
2\!\!\!\sum_{\substack{i<j<k\\j+k\text{ odd}\\i+k\text{ odd}}} a_i a_k {a_j}^2 e^{-x_{ik}} 
-2\!\!\!\sum_{\substack{i<j<k\\j+k\text{ even}\\i+k\text{ odd}}} a_i a_k {a_j}^2 e^{-x_{ik}} 
=
2\!\!\!\sum_{\substack{i<j<k\\ j+k\text{ odd}\\i+j\text{ even}}} a_i a_k {a_j}^2 e^{-x_{ik}} 
-2\!\!\!\sum_{\substack{i<j<k\\ j+k\text{ even}\\i+j\text{ odd}}} a_i a_k {a_j}^2 e^{-x_{ik}} 
\end{equation}
These two triple sums cancel the two triple sums \eqref{I-3sum3}. 
Hence, all sums have canceled, and therefore $\dot I_{(N)}=0$. 

This proof can be carried through for the unordered dynamical system \eqref{Npeakonsys}
by use of the sign identities
$\sgn(x-y)\sgn(z-x) + \sgn(y-z)\sgn(x-y) + \sgn(z-x)\sgn(y-z) = -1$
and 
$\sgn(w-x)\sgn(w-y)\sgn(w-z) +\sgn(x-w)\sgn(x-y)\sgn(x-z) +\sgn(y-w)\sgn(y-x)\sgn(y-z) 
+ \sgn(z-w)\sgn(z-x)\sgn(z-y) =0$, 
with $w,x,y,z\in\Rnum$.

\subsection{Proof of Theorem~\ref{Hamilstruct-Neven}}

First, consider the terms
\begin{equation}\label{eom-1stsum}
-3\sum_{k\neq j} \sgn(x_{ij}) \sgn(x_{jk}) a_k a_j e^{-|x_{jk}|} 
= 3\sum_{j<k} ( \sgn(x_{ik}) -\sgn(x_{ij}) ) a_j a_k e^{-x_{jk}} 
\end{equation}
and decompose this sum into the parts $i=j<k$, $j<k=i$, $j<i<k$, $i<j<k$, $j<k<i$,
where the index $i$ is fixed. 
The first two parts yield single sums
\begin{equation}\label{eom-1stsum1}
3\sum_{k>i} a_k a_i e^{-x_{ik}} + 3\sum_{j<i} a_j a_i e^{-x_{ji}} 
= 3a_i \sum_{j\neq i} a_j e^{-|x_{ij}|} , 
\end{equation}
and third part yields a double sum 
\begin{equation}\label{eom-1stsum2}
6\sum_{j<i<k} a_j a_k e^{-x_{jk}} , 
\end{equation}
while the last two parts vanish. 
Hence, this yields 
\begin{equation}\label{eom-1stsum-allterms}
3\sum_{j<k} ( \sgn(x_{ik}) -\sgn(x_{ij}) ) a_j a_k e^{-x_{jk}} 
= 3\Big( a_i \sum_{j\neq i} a_j e^{-|x_{ij}|} + 2\sum_{j<i<k} a_j a_k e^{-x_{jk}} \Big) . 
\end{equation}

Next, consider the remaining terms
\begin{equation}\label{eom-2ndsum}
\sum_{k\neq j} \sgn(x_{ij}) \sgn(x_{jk}) (-1)^{s_j + s_k} {a_k}^2 
=\sum_{j<k} (-1)^{j +k} ( \sgn(x_{ij}) {a_k}^2 - \sgn(x_{ik}) {a_j}^2 )
\end{equation}
and decompose this sum in the same way as the previous sum. 
The two parts $i=j<k$ and $j<k=i$ combine to yield
\begin{equation}\label{eom-2ndsum1}
-{a_i}^2 \Big(\sum_{l<i} (-1)^{i +l} + \sum_{l>i} (-1)^{i +l}\Big)
= {a_i}^2 S_i, 
\quad
S_i = -\sum_{l\neq i} (-1)^{i +l} . 
\end{equation}
The part $j<i<k$ can be split into two sums
which respectively combine with the sums given by the remaining two parts $i<j<k$ and $j<k<i$. 
This yields 
\begin{align}
& \sum_{i<j<k} (-1)^{j +k} ( {a_k}^2 - {a_j}^2 ) 
-\sum_{j<i<k} (-1)^{j +k} {a_k}^2 
= \sum_{l>i} (-1)^l {a_l}^2  S_{i,l} , 
\label{eom-2ndsum2}
\\
& 
\sum_{j<k<i} (-1)^{j +k} ( {a_j}^2 - {a_k}^2 )
-\sum_{j<i<k} (-1)^{j +k} {a_j}^2
= \sum_{l<i} (-1)^l {a_l}^2 S_{l,i} , 
\label{eom-2ndsum3}
\end{align}
where
\begin{equation}
S_{i,l} = \sum_{i<j<l}(-1)^j - \sum_{j<i}(-1)^j - \sum_{j>l}(-1)^j . 
\end{equation}
By use of the identity $2\sum_{j=1}^{n} (-1)^j = (-1)^n -1$, 
it is easy to evaluate $S_i$ and $S_{i,l}$ by decomposing the sums into separate cases 
where $i,l,N$ are even and odd:
\begin{equation}
S_i 
= \begin{cases}
1, & N \text{ even}
\\
0 , & N \text{ odd}, i \text{ odd}
\\
2 , & N \text{ odd}, i \text{ even}
\end{cases} , 
\quad
S_{i,l}
= \begin{cases}
0, & N \text{ even}
\\
1 , & N \text{ odd}
\end{cases} . 
\end{equation}
The combined terms \eqref{eom-2ndsum1}--\eqref{eom-2ndsum3} 
then yield 
${a_i}^2$ when $N$ is even,
$\sum_{l\neq i} (-1)^l {a_l}^2$ when $N$ is odd and $i$ is odd, 
and $\sum_{l\neq i} (-1)^l {a_l}^2 +2{a_i}^2$ when $N$ is odd and $i$ is even. 
These cases can be merged to give 
\begin{equation}\label{eom-2ndsum-allterms}
\sum_{j<k} (-1)^{j +k} ( \sgn(x_{ij}) {a_k}^2 - \sgn(x_{ik}) {a_j}^2 )
= {a_i}^2  +\tfrac{1}{2}(1-(-1)^N)\sum_{l\neq i} (-1)^l {a_l}^2  . 
\end{equation}
Finally,
adding these terms \eqref{eom-2ndsum-allterms} to the previous terms \eqref{eom-1stsum-allterms}
and substituting them into the Poisson bracket \eqref{NpeakFI-eomx}
yields the result 
\begin{equation}
\{ x_i, I_{(N)} \} = 
{a_i}^2  +\tfrac{1}{2}(1-(-1)^N)\sum_{l\neq i} (-1)^l {a_l}^2  
+ 3\Big( a_i \sum_{j\neq i} a_j e^{-|x_{ij}|} + 2\sum_{j<i<k} a_j a_k e^{-x_{jk}} \Big) . 
\end{equation}
This expression agrees with the $N$-peakon 
equation of motion \eqref{orderedNpeakonsys} for $x_i$
if and only if $N$ is even and the normalization factor is $\tfrac{\lambda}{2}=\tfrac{2}{3}$.

\end{document}